\documentstyle[epsf]{article}

\newcommand{\mib}[1]{\mbox{\boldmath $#1$}}
\newcommand{\mibs}[1]{\mbox{\scriptsize{\boldmath$#1$}}}

\newcommand{\pr}[3]{Phys.\ Rev. {\bf #1} (#2) #3}
\newcommand{\prb}[3]{Phys.\ Rev.\ B {\bf #1} (#2) #3}
\newcommand{\prl}[3]{Phys.\ Rev.\ Lett. {\bf #1} (#2) #3}
\newcommand{\rmp}[3]{Rev.\ Mod.\ Phys. {\bf #1} (#2) #3}
\newcommand{\jpsj}[3]{J.\ Phys.\ Soc.\ Jpn. {\bf #1} (#2) #3}
\newcommand{\ptp}[3]{Prog.\ Theor.\ Phys. {\bf #1} (#2) #3}

\newcommand{\epjb}[3]{Eur.\ Phys.\ J. B {\bf #1} (#2) #3}
\newcommand{\jpf}[3]{J.\ Phys.\ (France) {\bf #1} (#2) #3}

\newcommand{\del}{\partial}
\newcommand{\Tr}{{\rm Tr}}
\newcommand{\Tc}{T_{\rm c}}

\newcommand{\rmc}{{\rm c}}
\newcommand{\rme}{{\rm e}}
\newcommand{\romap}{{\rm p}}
\newcommand{\rms}{{\rm s}}

\newcommand{\calL}{{\mathcal L}}

\newcommand{\veps}{\varepsilon}
\newcommand{\bfveps}{\mib{\veps}}
\newcommand{\vepsz}{\veps^{(0)}}
\newcommand{\vepskin}{\veps_{\rm kin}}

\newcommand{\vepscor}[1]{\veps_{{\rm cor}#1}}
\newcommand{\bfphi}{\mib{\phi}}
\newcommand{\bfvarphi}{\mib{\varphi}}
\newcommand{\Ekin}{E_{\rm kin}}

\newcommand{\bfchi}{\mib{\chi}}

\newcommand{\muz}{\mu^{(0)}}
\newcommand{\bfSigma}{\mib{\Sigma}}

\newcommand{\ho}{{\hat{\omega}}}
\newcommand{\delho}{\delta\ho}
\newcommand{\ave}[1]{{\langle#1\rangle}}
\newcommand{\com}[2]{\left[#2\right]_{#1}\!}
\newcommand{\up}{\uparrow}
\newcommand{\down}{\downarrow}

\newcommand{\Psid}{\Psi^\dagger}

\newcommand{\bfPsi}{{\bf\Psi}}
\newcommand{\bfPsid}{\bfPsi^\dagger}

\newcommand{\BSCCO}[5]{Bi${}_{#1}$Sr${}_{#2}$Ca${}_{#3}$Cu${}_{#4}$O${}_{#5}$}

\title{Operator Projection Theory for Electron Differentiation in Underdoped Cuprate Superconductors\footnote{This is a proceeding of an invited talk presented at the international conference on Spectroscopies of Novel Superconductors 2001 at Chicago.}
}

\author{Shigeki Onoda\thanks{E-mail address: onoda@issp.u-tokyo.ac.jp},
Masatoshi Imada\thanks{E-mail address: imada@issp.u-tokyo.ac.jp}}

\date{\it Institute for Solid State Physics, University of Tokyo,
Kashiwanoha 5-1-5, Kashiwa, Chiba 277-8581, Japan}

\begin{document}
\maketitle
\begin{abstract}
Metals approaching the Mott insulator generate a new hierarchy in the electronic structure accompanied by a momentum dependent electron differentiation, beyond the Mott-Hubbard, Brinkman-Rice and Slater pictures of the Mott transition. To consider such nonlinear phenomenon, we develop an analytic nonperturbative theory based on operator projections combined with a self-consistent treatment of the low-energy excitations. This reproduces the formation of the Hubbard bands, Mott gap, spin fluctuations, mass divergence, diverging charge compressibility, and strongly renormalized flat and damped dispersion similar to angle-resolved photoemission data in high-$\Tc$ cuprates. Main structures in electronic spectra show a remarkable similarity to numerical results.
\end{abstract}

\noindent
{\it Keywords}: metal-insulator transition; projection operator; high-$\Tc$ cuprate superconductors

\vspace*{10pt}

\section{Introduction}\label{sec:Introduction}

The issues of metal-insulator transitions (MIT)~\cite{Mott} and the high-$\Tc$ cuprate superconductivity~\cite{BednortzMuller86} have opened extensive studies on related subjects from both theoretical and experimental aspects~\cite{RMP_Imada,Hubbard,BrinkmanRice70,Fulde}. In spite of a long history of studies~\cite{RMP_Imada,Fulde,Moriya_Spin-Fluctuations,DMFT}, their appropriate and satisfactory theoretical descriptions remain controversial. This paper aims at giving a unified view of the filling-control MIT in the 2D Hubbard model by means of the operator projection method (OPM)~\cite{OPM} and discussing {\it electron differentiation in the momentum space} obtained by the OPM which also occurs in underdoped high-$\Tc$ cuprates. The OPM is a recently developed non-perturbative analytic method based on the projeciton technique in an operator space~\cite{OP,Fulde} and systematically improves the Hartree-Fock theory and the Hubbard~\cite{Hubbard} and two-pole approximations~\cite{HarrisLange67,Roth69} as well as the perturbation and spin fluctuation theories~\cite{KadanoffBaym,FLEX,TPSC,DeiszHessSerene96,SCRPA}.

Pictures of early theories on the MIT are classified into three types: (i) The Hubbard approximations described an MIT as the splitting of two Hubbard bands due to the strong local Coulomb repulsion $U$ compared with the bare bandwidth $W$~\cite{Hubbard}. However, they and improved theories~\cite{HarrisLange67,Roth69} violate the Luttinger sum rule in paramagnetic metals. (ii) Brinkman and Rice predicted that at half filling, an MIT occurs at $U_{\rm c}(>0)$ due to the disappearance of quasiparticles accompanied by the mass divergence~\cite{BrinkmanRice70}. However, it ignores spin correlations and incoherent Hubbard bands. (iii) The Slater gives a view that a folding of the Brillouin zone accompanied by the antiferromagnetic (AF) phase transition yields an energy gap in the single-particle excitations. Self-consistent renormalization theory modified the Hartree-Fock theory and the RPA~\cite{Moriya_Spin-Fluctuations}. However, they do not give Mott-insulating features above the Neel temperature.

Further intensive studies on the MIT have been promoted to explain peculiar features observed in high-$\Tc$ cuprates~\cite{BednortzMuller86,RMP_Imada,Spingap,ARPES}. Their proper understanding and description remains still challenging due to difficulties in correctly describing strong correlations near the MIT.
It is crucial to build a unified theory of the three aspects of the MIT, in order to explain global phase diagrams of compounds that undergo an MIT~\cite{Mott,RMP_Imada}, to clarify the mechanism of the high-$\Tc$ superconductivity and to describe the MIT in the Hubbard model.

Gaussian fluctuations around the slave-boson mean-field theory~\cite{SBMF} as well as the dynamical mean-field solution (DMFT)~\cite{DMFT} incorporates the Hubbard bands but not spin correlations to the Brinkman-Rice picture.  The DMFT predicts a lowest-energy resonance inside the Hubbard gap at half filling below $U_{\rm c}(>0)$. However, it ignores the momentum dependence of the self-energy part in finite dimensions. Some independent momentum patches can be inctroduced into this local self-energy part~\cite{DCA}. Then, for the 2D half-filled Hubbard model at weak couplings, a Mott insulator with sharp peaks at the gap edges is reproduced, though solving a cluster problem yields severe limitations here.

No theories self-consistently modifying the RPA~\cite{KadanoffBaym,FLEX,TPSC,DeiszHessSerene96,SCRPA} predict Mott-insulating features~\cite{DeiszHessSerene96}, without abandoning the self-consistency between the Green's function and the self-energy part~\cite{TPSC}.

Quantum Monte-Carlo (QMC) simulations have revealed the thermodynamic and dynamical properties of the 2D Hubbard model: The charge compressibility diverges towards half filling~\cite{FurukawaImada}. Near half filling, nearly separated Hubbard bands develop, the Luttinger sum rule is violated at the temperature $T\ge J=4t^2/U$ and a sharp peak of the local density of states develops near the Fermi level with decreasing $T$, reflecting a suppressed Fermi temperature $T_{\rm F}$.~\cite{BulutScalapinoWhite94}. At $T<J$, extra two bands appear in addition to the Hubbard bands~\cite{HankePreuss95,Grober}. A dispersion around the $(\pi,0)$ and $(0,\pi)$ momenta is strongly weak~\cite{Assaad}. Strong momentum dependence of quasiparticle excitations is a crucial feature beyond the early three pictures of the MIT.

The OPM~\cite{OPM} offers a theoretical framework unifying the early three pictures and reproducing key features beyond them~\cite{FurukawaImada,BulutScalapinoWhite94,HankePreuss95,Grober,Assaad}, the momentum-dependent electron differentiation in the low-energy hierarchy. They are discussed in \S~\ref{sec:half-filling} and \ref{sec:Doped}.

The OPM is formulated in \S~\ref{sec:Formulation}. Based on equations of motion, it gives a unique continued-fraction or Dyson-equation expression of a correlation function. It may systematically extract the lower-energy properties by projecting out higher-energy degrees of freedom in a self-consistent fashion. For the self-energy part, we can build its Dyson equation beyond the perturbation and one-loop theories. The higher-order dynamics was considered within a two-site level in a local picture~\cite{MatsumotoMancini97}, though growing spin correlations were ignored even near the continuous AF phase transition. The OPM solves this difficulty~\cite{OPM}.

In \S~\ref{sec:ARPES}, we discuss the similarity between our results and observed normal-state properties in high-$\Tc$ cuprates.

\S~\ref{sec:Summary} is devoted to the summary of this paper.

\section{Operator projection method}
\label{sec:Formulation}

We follow the previous papers~\cite{OPM}.
We consider the Hubbard Hamiltonian with an electron transfer $t_{\mibs{x},\mibs{x}'}(=t_{\mibs{x}',\mibs{x}})$ from an atomic site $\mib{x}'$ to $\mib{x}$;
$H\equiv-\sum_{\mibs{x},\mibs{x}',s}t_{\mibs{x},\mibs{x}'}c^\dagger_{\mibs{x}s}c_{\mibs{x}'s}+U\sum_{\mibs{x}}n_{\mibs{x}\up}n_{\mibs{x}\down}$, with $n_{\mibs{x}s}\equiv c^\dagger_{\mibs{x}s}c_{\mibs{x}s}$.
We perform the projection procedure for the electron creation and annihilation operators at a site $\mib{x}$ with a spin index $s$, $c^\dagger_{\mibs{x}s}$ and $c_{\mibs{x}s}$.
It is useful to define the fermionic $4N$-component vector operator $\bfPsi$ and its Hermitian conjugate $\bfPsid$, composed of
\begin{eqnarray}
\Psi_{\mibs{x}}&=&{}^t(c_{\mibs{x}\up},c_{\mibs{x}\down},-c^\dagger_{\mibs{x}\down},c^\dagger_{\mibs{x}\up}),
\\
\Psi^\dagger_{\mibs{x}}&=&(c^\dagger_{\mibs{x}\up},c^\dagger_{\mibs{x}\down},-c_{\mibs{x}\down},c_{\mibs{x}\up}),
\end{eqnarray}
respectively, in the real-space representation with the number of the atoms $N$. We define the commutator of an operator $A$ with $\bar{H}\equiv H-\mu\sum_{\mibs{x}s}n_{\mibs{x}s}$ as $\ho A\equiv[A,\bar{H}]_-$ and the thermal average of $A$ at temperature $T$ as $\ave{A}\equiv\Tr[e^{-\bar{H}/T}A]/Z$ with the partition function $Z\equiv\Tr[e^{-\bar{H}/T}]$ and the chemical potential $\mu$. The Heisenberg operator is introduced as $A(t)\equiv e^{i\bar{H}t}A e^{-i\bar{H}t}$. Then, for arbitrary $4N$-component vector operators with an odd number of particles $\bfphi$ and $\bfvarphi$, their response function and susceptibility are defined in the $4N\times4N$ matrix representation as $\mib{K}_{\phi,\varphi^\dagger}(t)$ and $\bfchi_{\phi,\varphi^\dagger}(\omega)=\int_0^\infty\!dt\,e^{i\omega t}\mib{K}_{\phi,\varphi^\dagger}(t)$, respectively. Here, the $(a,a')\otimes(\mib{x},\mib{x}')$ component of $\mib{K}_{\phi,\varphi^\dagger}(t)$ is defined by
\begin{equation}
K^{aa'}_{\phi,\varphi^\dagger}(t)\equiv-i\ave{\com{+}{\phi^a_{\mibs{x}}(t),(\varphi^\dagger){}^{a'}_{\mibs{x}'}}}.
\end{equation}

The first-order projection is given by
\begin{eqnarray}
\ho\bfPsi&=&\bfveps^{(11)}\bfPsi+\delho\bfPsi,
\\
\bfveps^{(11)}&\equiv&\mib{K}_{\ho\Psi,\Psid}(0)\mib{K}^{-1}_{\Psi,\Psid}(0),
\\
\delho\bfPsi&\equiv&(1-P_1)\ho\bfPsi.
\end{eqnarray}
$P_1$ is defined as the first-order projection operator which operates to an arbitrary operator $A$ as $P_1A=\mib{K}_{A,\Psid}(0)\mib{K}^{-1}_{\Psi,\Psid}(0)\bfPsi$ and satisfies $P_1^2=P_1$.
$\delho\bfPsi$ satisfies $\mib{K}_{\Psi,(\delho\Psi)^\dagger}(0)=\mib{K}_{\delho\Psi,\Psid}(0)=0$.
This leads to the Dyson equation for the single-particle Green's function $\mib{G}(\omega)=\bfchi_{\Psi,\Psid}(\omega)$
in the $4N\times4N$ matrix representation,
\begin{eqnarray}
\mib{G}(\omega)&=&\big[\big(\omega\mib{I}-\bfveps^{(11)}\big)-\bfSigma_1(\omega)\big]^{-1},
\label{eq:1OP:Dyson-G}\\
\bfSigma_1(\omega)&=&-i\bfchi_{\delho\Psi,(\delho\Psi)^\dagger}^{\rm irr}(\omega)\mib{K}_{\Psi,\Psid}^{-1}(0).
\label{eq:1OP:Dyson-chi_Psi}
\end{eqnarray}
$\mib{I}$ is the $4N\times4N$ identity matrix.
$\bfchi_{\delho\Psi,(\delho\Psi)^\dagger}^{\rm irr}(\omega)
 =[\bfchi_{\delho\Psi,(\delho\Psi)^\dagger}^{-1}(\omega)-i\mib{K}_{\Psi,\Psid}^{-1}(0)\mib{G}^{(0)}(\omega)]^{-1}$ is the irreducible part of $\bfchi_{\delho\Psi,(\delho\Psi)^\dagger}(\omega)$ with respect to $\mib{G}^{(0)}(\omega)$.
We define the local charge, spin and pair operators at $\mib{x}$ as
$n_{\mibs{x}}\equiv n_{\mibs{x}\up}+n_{\mibs{x}\down}$,
$\mib{S}_{\mibs{x}}\equiv c^\dagger_{\mibs{x}s}\mib{\sigma}_{ss'}c_{\mibs{x}s'}/2$,
$\Delta^{1}_{\mibs{x}}\equiv\frac{1}{\sqrt{2}}(c_{\mibs{x}\up}c_{\mibs{x}\down}+c^\dagger_{\mibs{x}\down}c^\dagger_{\mibs{x}\up})$ and
$\Delta^{2}_{\mibs{x}}\equiv\frac{i}{\sqrt{2}}(c_{\mibs{x}\up}c_{\mibs{x}\down}-c^\dagger_{\mibs{x}\down}c^\dagger_{\mibs{x}\up})$.
We introduce the identity and Pauli matrices $\sigma_0$ ($\rho_0$) and $\mib{\sigma}$ ($\mib{\rho}$), respectively, which operate to the spin (charge) space of vector operators. Then, for $U\ge0$, one obtains
\begin{eqnarray}
\veps^{(11)}_{\mibs{x},\mibs{x}'}
&=&\vepsz_{\mibs{x},\mibs{x}'}\rho_3\sigma_0
-U\ave{S^z_{\mibs{x}}}\delta_{\mibs{x},\mibs{x}'}\rho_0\sigma_3,
\label{eq:1OP:veps11_x}\\
\vepsz_{\mibs{x},\mibs{x}'}&=&-t_{\mibs{x},\mibs{x}'}-(\mu-U\ave{n_{\mibs{x}}}/2)\delta_{\mibs{x},\mibs{x}'},
\\
\delho\Psi_{\mibs{x}}&=&U\left(\delta n_{\mibs{x}}\rho_3\sigma_0/2-\delta S^z_{\mibs{x}}\rho_0\sigma_3\right)\Psi_{\mibs{x}},
\label{eq:1OP:delhho-Psi_x}
\end{eqnarray}
with $\delta n_{\mibs{x}}=n_{\mibs{x}}-\ave{n_{\mibs{x}}}$ and
$\delta S^z_{\mibs{x}}=S^z_{\mibs{x}}-\ave{S^z_{\mibs{x}}}$.
We have taken $\ave{\Delta^{i}_{\mibs{x}}}=0$, and also $\ave{S^{\pm}_{\mibs{x}}}=0$ without loss of generality.

The present theory without $\bfSigma_1(\omega)$ yields the Hartree-Fock theory. Self-consistent calculations of $\bfSigma_1(\omega)$ by perturbation or one-loop theories~\cite{KadanoffBaym,FLEX} do not reproduce the Mott insulator~\cite{TPSC,DeiszHessSerene96}.

Next, we construct the Dyson equation for $\bfSigma_1(\omega)$, performing the second-order projection;
\begin{eqnarray}
&&\hspace*{-15pt}\ho\delho\bfPsi=\bfveps^{(21)}\bfPsi+\bfveps^{(22)}\delho\bfPsi+\delho\delho\bfPsi,
\\
&&\hspace*{-10pt}\bfveps^{(21)}\equiv\mib{K}_{\ho\delho\Psi,\Psid}(0)\mib{K}^{-1}_{\Psi,\Psid}(0),
\\
&&\hspace*{-10pt}\bfveps^{(22)}\equiv\mib{K}_{\ho\delho\Psi,(\delho\Psi)^\dagger}(0)
\mib{K}^{-1}_{\delho\Psi,(\delho\Psi)^\dagger}(0),
\\
&&\hspace*{-22pt}\delho\delho\bfPsi\equiv(1-P_2)\ho\delho\bfPsi.
\end{eqnarray}
$P_2$ is defined as the second-order projection operator which operates to an arbitrary operator $A$ as
$P_2A=P_1A+\mib{K}_{A,(\delho\Psi)^\dagger}(0)\mib{K}^{-1}_{\delho\Psi,(\delho\Psi)^\dagger}(0)\delho\bfPsi$ and satisfies $P_2^2=P_2$. Then, we obtain
\begin{eqnarray}
\hspace*{-20pt}\bfSigma_1(\omega)&=&\big[\big(\omega\mib{I}-\bfveps^{(22)}\big)-\bfSigma_2(\omega)\big]^{-1}\bfveps^{(21)},
\label{eq:2OP:Sigma_1}\\
\hspace*{-20pt}\bfSigma_2(\omega)
&=&-i\bfchi^{\rm irr}_{\delho\delho\Psi,(\delho\delho\Psi)^\dagger}(\omega)
\mib{K}^{-1}_{\delho\Psi,(\delho\Psi)^\dagger}(0).
\label{eq:2OP:Sigma_2}
\end{eqnarray}
Here, $\bfchi_{\delho\delho\Psi,(\delho\delho\Psi)^\dagger}^{\rm irr}(\omega)=[\bfchi^{-1}_{\delho\delho\Psi,(\delho\delho\Psi)^\dagger}(\omega)-i\mib{K}^{-1}_{\delho\Psi,(\delho\Psi)^\dagger}(0)\bfSigma_1^{(0)}(\omega)]^{-1}$ is the irreducible part of $\bfchi_{\delho\delho\Psi,(\delho\delho\Psi)^\dagger}(\omega)$ with respect to $\mib{G}^{(0)}(\omega)$ and $\bfSigma_1(\omega)$. For the repulsive Hubbard model,
\begin{eqnarray}
&&\hspace*{-18.5pt}
\veps^{(21)}_{\mibs{x},\mibs{x}'}=U^2\delta_{\mibs{x},\mibs{x}'}M_{\mibs{x}},
\label{eq:2OP:veps21_x}\\
&&\hspace*{-12pt}
M_{\mibs{x}}\equiv\ave{\left(\delta n_{\mibs{x}}\rho_3\sigma_0/2-\delta S^z_{\mibs{x}}\rho_0\sigma_3\right)^2}
\\
&&\hspace*{-18.5pt}
\veps^{(22)}_{\mibs{x},\mibs{x}'}=-t^{(22)}_{\mibs{x},\mibs{x}'}\rho_3\sigma_0-\left(\mu_2\rho_3\sigma_0-\ave{S^z_x}\rho_0\sigma_3\right)
\delta_{\mibs{x},\mibs{x}'},
\label{eq:2OP:veps22_x}\\
&&\hspace*{-18pt}
t^{(22)}_{\mibs{x},\mibs{x}'}\equiv t_{\mibs{x},\mibs{x}'}
\langle\delta\calL_{\mibs{x}}\rho_3\delta\calL_{\mibs{x}'}\rho_3\rangle M_{\mibs{x}'}^{-1},
\label{eq:2OP:t^2_x}\\
&&\hspace*{-14.5pt}
\delta\calL_{\mibs{x}}\equiv\delta n_{\mibs{x}}\rho_3\sigma_0/2
-\delta\mib{S}_{\mibs{x}}\cdot\rho_0\mib{\sigma}
+\Delta^{\bar{i}}_{\mibs{x}}\rho_{\bar{i}}\sigma_0/\sqrt{2},
\label{eq:2OP:delta-calL_x}\\
&&\hspace*{-8pt}
\mu_2=\muz_2\rho_0\sigma_0+\left[(1-\ave{n})\vepskin+\vepscor{2}\right]M_{\mibs{x}}^{-1},
\label{eq:2OP:mu2}\\
&&\hspace*{-13pt}
\vepskin=\frac{-1}{2N}t_{\bar{\mibs{k}}}(\ave{c^\dagger_{\bar{\mibs{k}}\bar{s}}c_{\bar{\mibs{k}}\bar{s}}}\rho_3\sigma_0-\ave{c^\dagger_{\bar{\mibs{k}}\bar{s}}\sigma_{3\bar{s}\bar{s}'}c_{\bar{\mibs{k}}\bar{s}'}}\rho_0\sigma_3),
\end{eqnarray}
\noindent
where $\muz_2\equiv\mu-U(1-\ave{n_{\mibs{x}}}/2)$. Hereafter, the summations over $\bar{i}=1$ and $2$ and barred variables as $\bar{\mib{x}}$ and $\bar{\mib{k}}$ should be taken.
Below, a local energy shift $\vepscor{2}$ due to two-site correlated hopping terms is neglected. We note that in the particle-hole symmetric case, $\vepscor{2}$ vanishes. Then, in the case of $\ave{S^z_{\mibs{x}}}=0$, we obtain
\begin{eqnarray}
\delho\delho\Psi_{\mibs{x}}&\approx&
\big(\delta_{\mibs{x},\bar{\mibs{x}}}(1-\ave{n})\vepskin M_{\mibs{x}}^{-1}
+t^{(22)}_{\mibs{x},\bar{\mibs{x}}}\big)\rho_3\delho\Psi_{\bar{\mibs{x}}}
\nonumber\\
&&{}-Ut_{\mibs{x},\bar{\mibs{x}}}\delta\calL_{\mibs{x}}\rho_3\Psi_{\bar{\mibs{x}}}.
\label{eq:2OP:delho-delho-Psi_x}
\end{eqnarray}

Without $\bfSigma_2(\omega)$ and the momentum dependence of $\bfveps^{(22)}$, this theory gives the Hubbard I approximation~\cite{Hubbard}. The momentum dependence of $\bfveps^{(22)}$ that includes a superexchange term modifies the dispersions of the Hubbard bands~\cite{Roth69}.

In the symmetry-unbroken phase, we have only to discuss the spin-independent electron part, $\Sigma_{2\rme}(\omega,\mib{k})=4\chi_{\delho\delho\Psi,(\delho\delho\Psi)^\dagger}^{{\rm irr} 11}(\omega,\mib{k})/\ave{n}(2-\ave{n})$. We take the following decoupling approximation to $\Sigma_{2\rme}(\omega,\mib{k})$ to correctly estimate its moment and to include the short-ranged correlations neglected in previous equation-of-motion approaches~\cite{Hubbard,Roth69,MatsumotoMancini97};
\begin{eqnarray}
\lefteqn{\Sigma_{2\rme}(\omega_n,\mib{k})=\frac{T/N}{\ave{n}(2\!-\!\ave{n})}\sum_{m,\mibs{q}}\Big[G(\omega_n\!-\!\Omega_m,\mib{k}\!-\!\mib{q})}
\nonumber\\
&&\hspace*{-5pt}\big\{T_{\mibs{k};\mibs{q}}^2
\left(\chi_\rmc(\Omega_m,\mib{q})\!+\!\chi_\rms(\Omega_m,\mib{q})\right)\!+\!2t_{\mibs{k}-\mibs{q}}^2\chi_\rms(\Omega_m,\mib{q})\!\big\}\!
\nonumber\\
&&-2t_{\mibs{k}-\mibs{q}}^2
G(\Omega_m\!-\!\omega_n,\mib{q}\!-\!\mib{k})\chi_\romap(\Omega_m,\mib{q})\Big],
\label{eq:2EQM-G:Sigma_2_decoupling}
\end{eqnarray}
where $T_{\mibs{k};\mibs{q}}=t_{\mibs{k}-\mibs{q}}-\frac{2(1-\ave{n})}{\ave{n}(2-\ave{n})}\Ekin-\tilde{t}_{\mibs{k}}$ with $\Ekin=\frac{-1}{N}\sum_{\mibs{k},s}t_{\mibs{k}}\ave{c^\dagger_{\mibs{k}s}c_{\mibs{k}s}}$ and $\tilde{t}_{\mibs{k}}$ is the $(1,1)$ component of $t^{(22)}_{\mibs{k}}$.
$\Omega_m=2\pi mT$ is a bosonic Matsubara frequency, $G_\rme$ is the spin-independent electron Green's function and $\chi_\rmc$, $\chi_\rms$ and $\chi_\romap$ are the charge, spin, local-pair susceptibilities, respectively.

We calculate the susceptibilities with the two-particle self-consistent method~\cite{TPSC} to include short-ranged correlations and to reproduce their local moments. Following Ref.~\cite{TPSC}, we take
\begin{eqnarray}
&&\hspace*{-20pt}\chi_{\rmc}(\Omega_m,\mib{k})\!=\!2\chi_{\rm ph}(\Omega_m,\mib{k})/[1\!+\!\Gamma_\rmc\chi_{\rm ph}(\Omega_m,\mib{k})],
\label{TPSC:chin}\\
&&\hspace*{-20pt}\chi_{\rms}(\Omega_m,\mib{k})\!=\!2\chi_{\rm ph}(\Omega_m,\mib{k})/[1\!-\!\Gamma_s\chi_{\rm ph}(\Omega_m,\mib{k})],
\label{TPSC:chis}\\
&&\hspace*{-20pt}\chi_{\romap}(\Omega_m,\mib{k})\!=\!\chi_{\rm pp}(\Omega_m,\mib{k})/[1\!+\!\Gamma_\romap\chi_{\rm pp}(\Omega_m,\mib{k})],
\label{TPSC:chip}
\end{eqnarray}
with an approximation $\Gamma_\rms=4U\ave{n_{\mibs{x}\up}n_{\mibs{x}\down}}/\ave{n}^2$ and the self-consistency conditions for $\Gamma_\rms$, $\Gamma_\rmc$ and $\Gamma_\romap$ given by the local moment sum rules;
$T/N\sum_{m,\mibs{k}}\chi_{n,n}(\Omega_m,\mib{k})=\ave{n}+2\ave{n_{\mibs{x}\up}n_{\mibs{x}\down}}-\ave{n}^2$,
$T/N\sum_{m,\mibs{k}}\chi_{S^i,S^i}(\Omega_m,\mib{k})
=\ave{n}-2\ave{n_{\mibs{x}\up}n_{\mibs{x}\down}}$,
and $T/N\sum_{m,\mibs{k}}\chi_{\Delta,\Delta^\dagger}(\Omega_m,\mib{k})
=\ave{n_{\mibs{x}\up}n_{\mibs{x}\down}}$.
$\chi_{\rm ph}$ and $\chi_{\rm pp}$ are the particle-hole and particle-particle susceptibilities calculated from the bare Green's functions.

Below, results are shown for the single-particle density of states $\rho(\omega)\equiv\frac{1}{N}\sum_{\mibs{k}}A(\omega, \mib{k})$, spectral functions $A(\omega,\mib{k})\equiv-\frac{1}{\pi}{\rm Im}G_\rme(\omega,\mib{k})$, dispersions and the $\mu$-$\ave{n}$ curve.

\section{Mott-insulating properties at half filling}
\label{sec:half-filling}

\begin{figure}[htb]
\begin{center}
\epsfysize=5.5cm
$$\epsffile{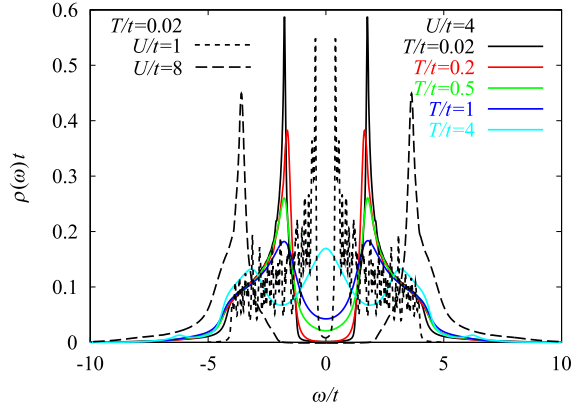}$$
\end{center}
\caption{The local density of states $\rho(\omega)$ at half filling $\ave{n}=1$. The Mott-Hubbard pseudogap remains even at $T=J\equiv 4t^2/U$ above which AFM spin correlations are not enhanced.}
\label{fig:DOSel1t0U1,4,8}
\end{figure}
\begin{figure}[htb]
\begin{center}
\epsfysize=5.5cm
$$\epsffile{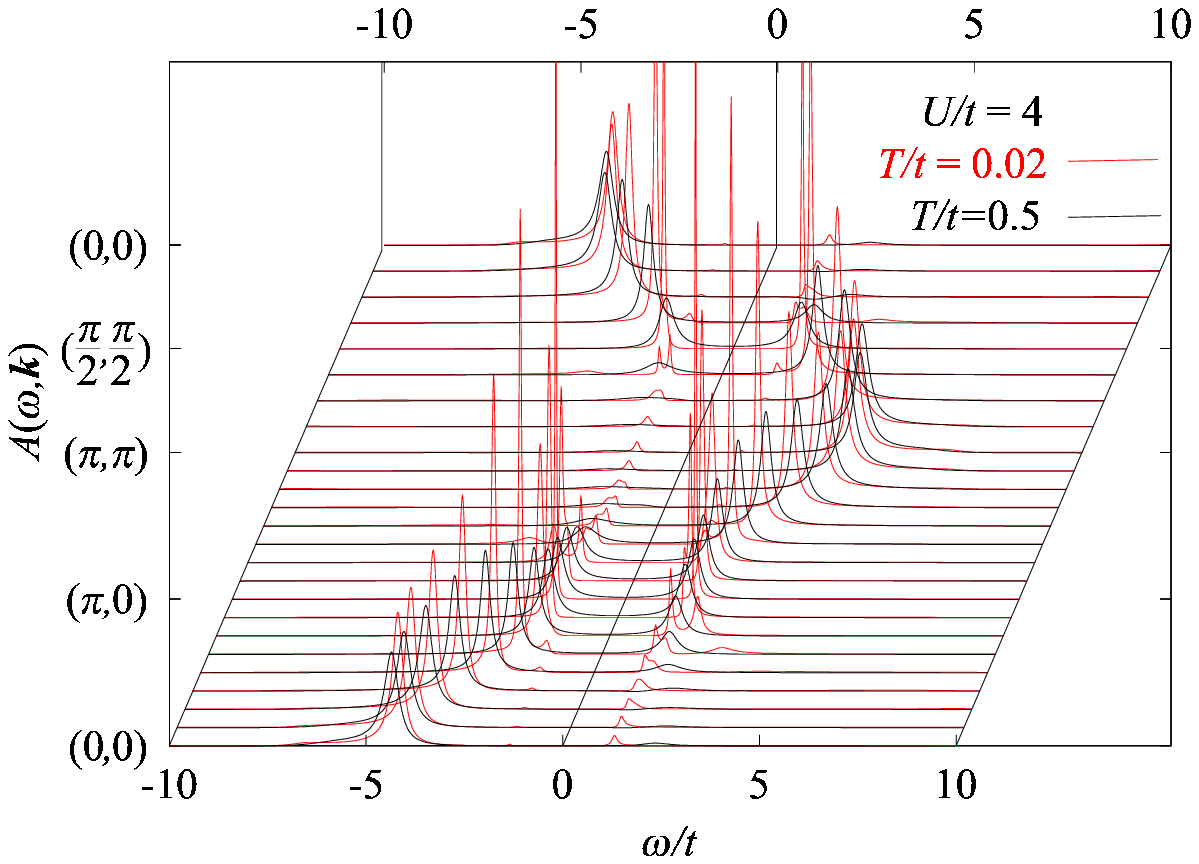}$$
\end{center}
\caption{$A(\omega,\mib{k})$ for $U/t=4$ in the case of half filling $\ave{n}=1$.}
\label{fig:Akwel1t0U4_T.02_.5_4}
\end{figure}
\begin{figure}[htb]
\begin{center}\leavevmode
\epsfysize=5.5cm
$$\epsffile{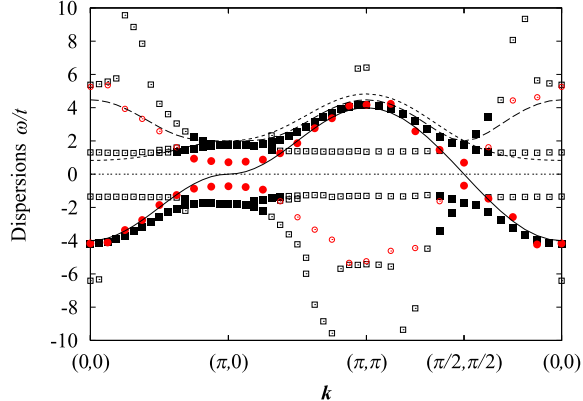}$$
\end{center}
\caption{Single-particle dispersions for $U/t=4$ and $T/t=0.02$ at $\ave{n}=1$. The squares and the red circles denote the present and QMC results~\cite{Assaad}, respectively. The open and the filled symbols represent the momenta where the peak intensities divided by the largest intensity are less and larger than $0.1$, respectively. The solid, dotted and dashed lines denote the free-electron dispersion and the upper Hubbard dispersion by the Hubbard I approximation and the two-pole approximation with $\tilde{t}_{\mibs{k}}=t_{\mibs{k}+\mibs{Q}}$, respectively.}
\label{fig:disel1t0U4_T.02_.5_4}
\end{figure}

Figure~\ref{fig:DOSel1t0U1,4,8} shows that at low temperatures, a Mott-insulating gap grows, indicating the Mott-insulating ground states. This is accompanied by sharp peaks of $\rho(\omega)$ at the low-energy edges of the Hubbard bands. This agrees with the QMC results~\cite{BulutScalapinoWhite94} but contrasts with the Hubbard approximations~\cite{Hubbard} and the DMFT~\cite{DMFT}. As shown in Figs.~\ref{fig:Akwel1t0U4_T.02_.5_4} and \ref{fig:disel1t0U4_T.02_.5_4}, the growth of AF spin fluctuations develops the AF shadows of the Hubbard bands, yielding a four-band structure, in agreement with QMC results~\cite{Grober}. The structure takes the form of a superposition of two SDW-like bands and two low-energy narrow bands at the gap edges. The low-energy bands have a particularly weak dispersion and dominant weights around $(\pi,0)$ and $(0,\pi)$. This is consistent with the QMC results~\cite{HankePreuss95,Grober,Assaad}. These excitation spectra around $(\pi,0)$ and $(0,\pi)$ mainly contribute to these sharp peaks. With increasing $T$, spectral weights fill in the gap and the intensities of the peaks are weakened. AF spin correlations also diminishes. This smears out the AF shadows. Then, the main structure is characterized by the Hubbard bands nearly separated by the Mott-Hubbard pseudogap, as shown in Fig.~\ref{fig:Akwel1t0U4_T.02_.5_4}.

For the momentum distribution, the error of the OPM compared with the $T=0$ QMC results turns out to be less than $10\%$ for $U/t=4$ and is even smaller for $U/t=8$~\cite{Assaad}. Moreover, Fig.~\ref{fig:disel1t0U4_T.02_.5_4} suggests that a main structure in $A(\omega,\mib{k})$ shows a remarkable similarity to the QMC results~\cite{Assaad}. This small discrepancy mainly originates from a small reduction of the Mott gap from $U$ compared with QMC data.

\section{Metallic properties in doped cases}
\label{sec:Doped}

\begin{figure}[htb]
\begin{center}\leavevmode
\epsfysize=5.5cm
$$\epsffile{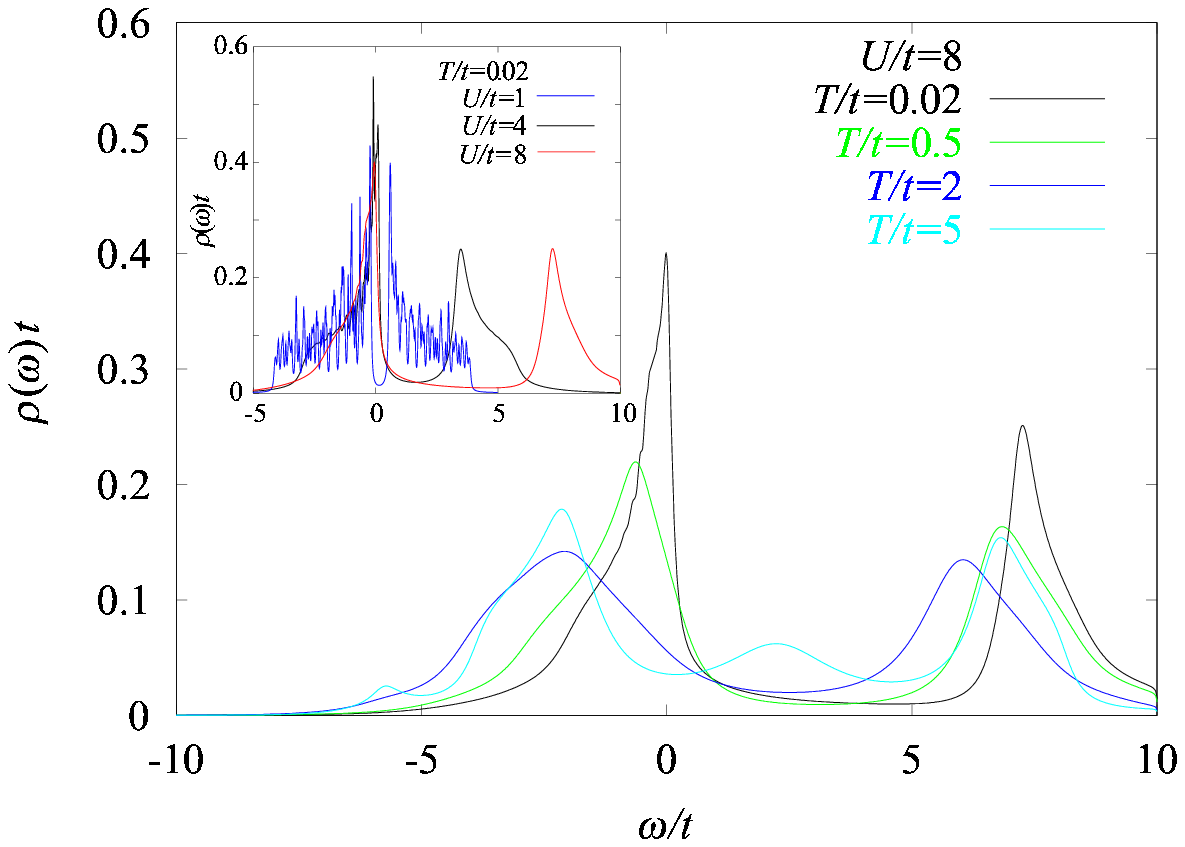}$$
\end{center}
\caption{The $T$ dependence of the local density of states $\rho(\omega)$ at $n=0.87$ for $U/t=8$. The inset shows the $U$ dependence at the same filling $n=0.87$ at $T/t=0.02$. A sharp peak develops near $\omega=0$ at low $T$.}
\label{fig:DOSel.87t0U1,4,8}
\end{figure}
\begin{figure}[htb]
\begin{center}\leavevmode
\epsfysize=5.5cm
$$\epsffile{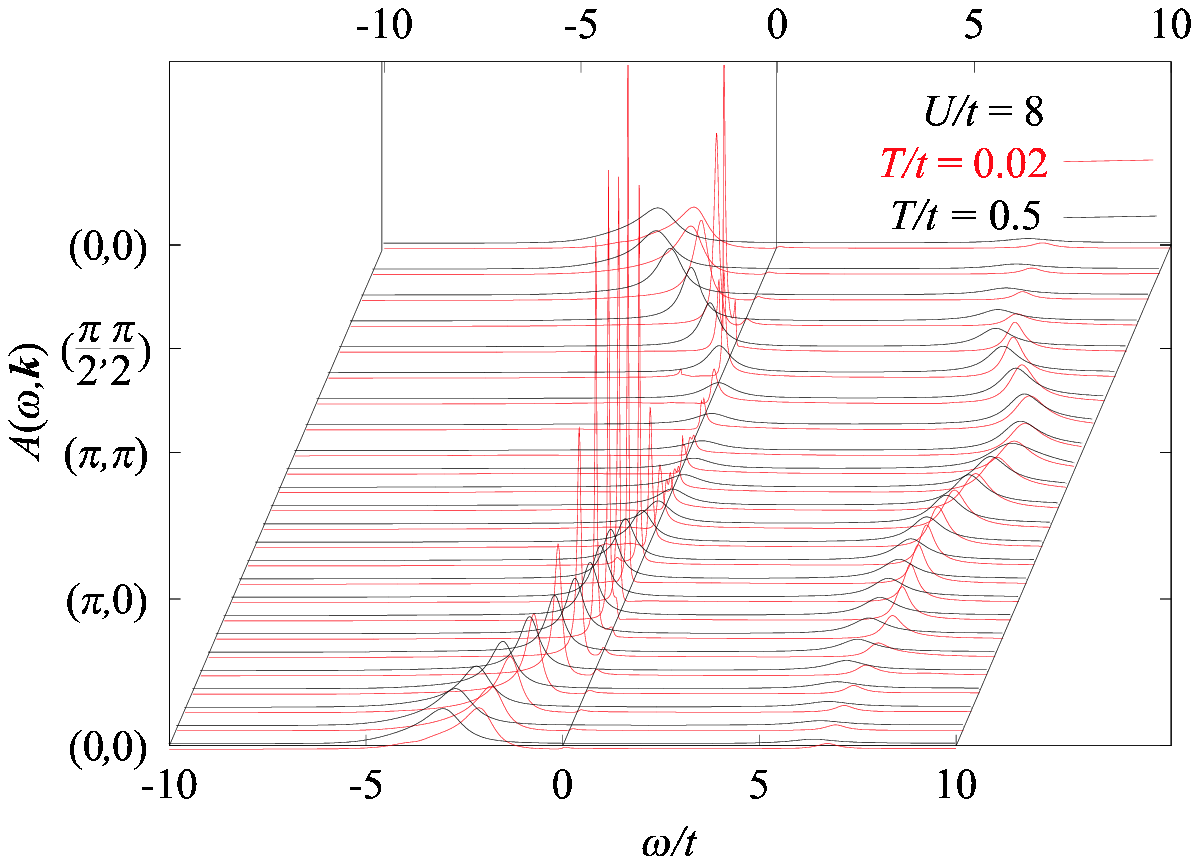}$$
\end{center}
\caption{$A(\omega,\mib{k})$ for $U/t=8$ and $\ave{n}=0.87$.}
\label{fig:Akwel.87t0U8_T.02_.5_5}
\end{figure}
\begin{figure}[htb]
\begin{center}\leavevmode
\epsfysize=5.5cm
$$\epsffile{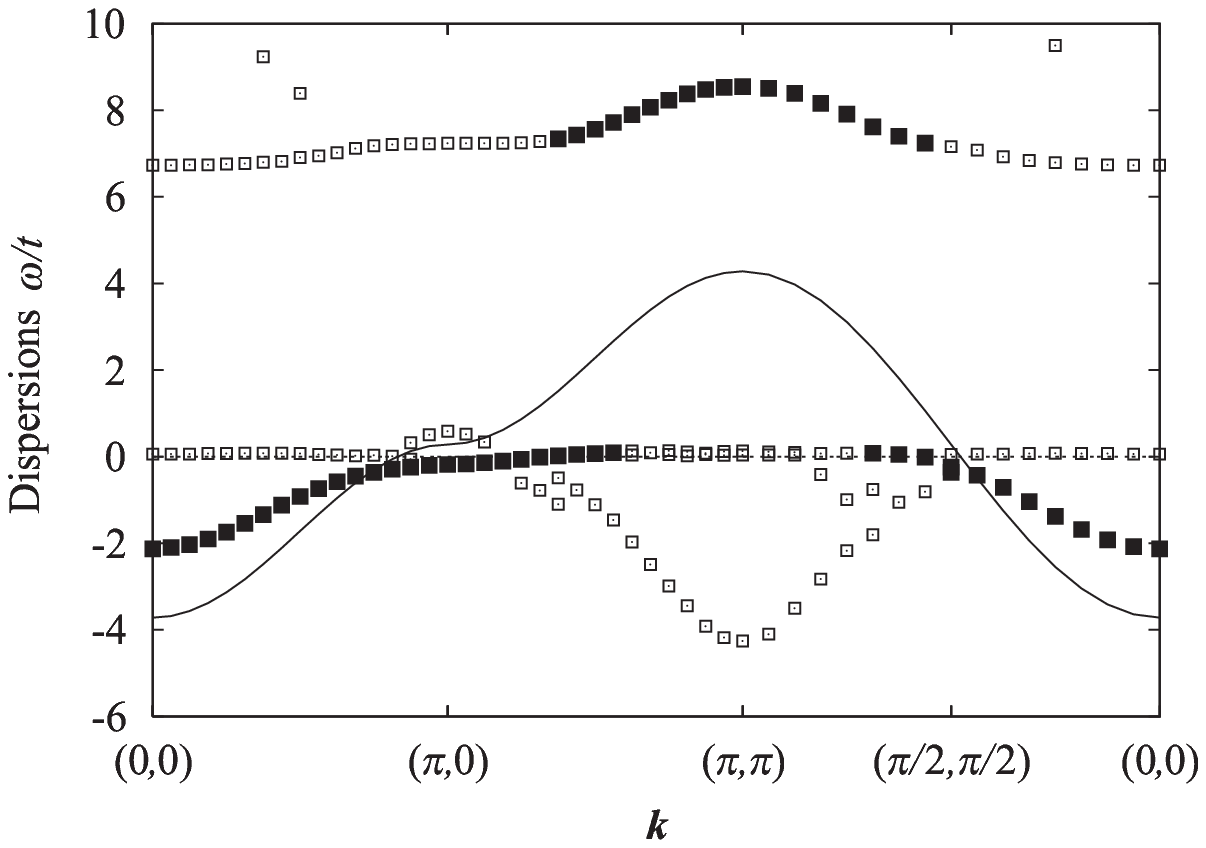}$$
\end{center}
\caption{Single-particle dispersions for $U/t=8$ and $\ave{n}=0.87$ at $T/t=0.02$ with the same symbols as in Fig.~\ref{fig:disel1t0U4_T.02_.5_4}.}
\label{fig:disel.87t0U8_T.02_.5_5}
\end{figure}

Results on $\rho(\omega)$ at $\ave{n}=0.87$ are shown in Fig.~\ref{fig:DOSel.87t0U1,4,8}. The upper and lower Hubbard bands are nearly separated by the Mott-Hubbard pseudogap with remaining weak incoherent spectra. This is also shown in Fig.~\ref{fig:Akwel.87t0U8_T.02_.5_5} that gives $A(\omega,\mib{k})$ similar to the results of the two-pole approximation~\cite{Roth69}. With decreasing $T$, a sharp peak of $\rho(\omega)$ develops near $\omega=0$. They agree with QMC results~\cite{BulutScalapinoWhite94}.
As shown in Figs.~\ref{fig:Akwel.87t0U8_T.02_.5_5} and \ref{fig:disel.87t0U8_T.02_.5_5}, the single-particle dispersion in the lower Hubbard band is strongly weak both outside the magnetic Brillouin zone boundary and around the $(\pi,0)$ and $(0,\pi)$ momenta at $T/t=0.02$, in agreement with QMC results~\cite{BulutScalapinoWhite94}.
Particularly, Fig.~\ref{fig:disel.87t0U8_T.02_.5_5} shows a formation of a low-energy narrow band {\it pinned} at the top of the lower Hubbard band. This band has an extremely weak dispersion and dominant weights around $(\pi,0)$ and $(0,\pi)$ and seems to evolve into the low-energy band obtained at half filling. This contributes to the sharp peak of the density of states at the top of the lower Hubbard band which persists even in the Mott insulator. This generates a divergence of the compressibility as $\ave{n}\to1$, as shown in Fig~\ref{fig:ef-elt0U1,4,8}.
 This contrasts with the Brinkman-Rice picture where the compressibility remains finite since the sharp peaks are absent in the Mott insulator~\cite{DMFT}.
\begin{figure}[htb]
\begin{center}
\epsfysize=5.5cm
$$\epsffile{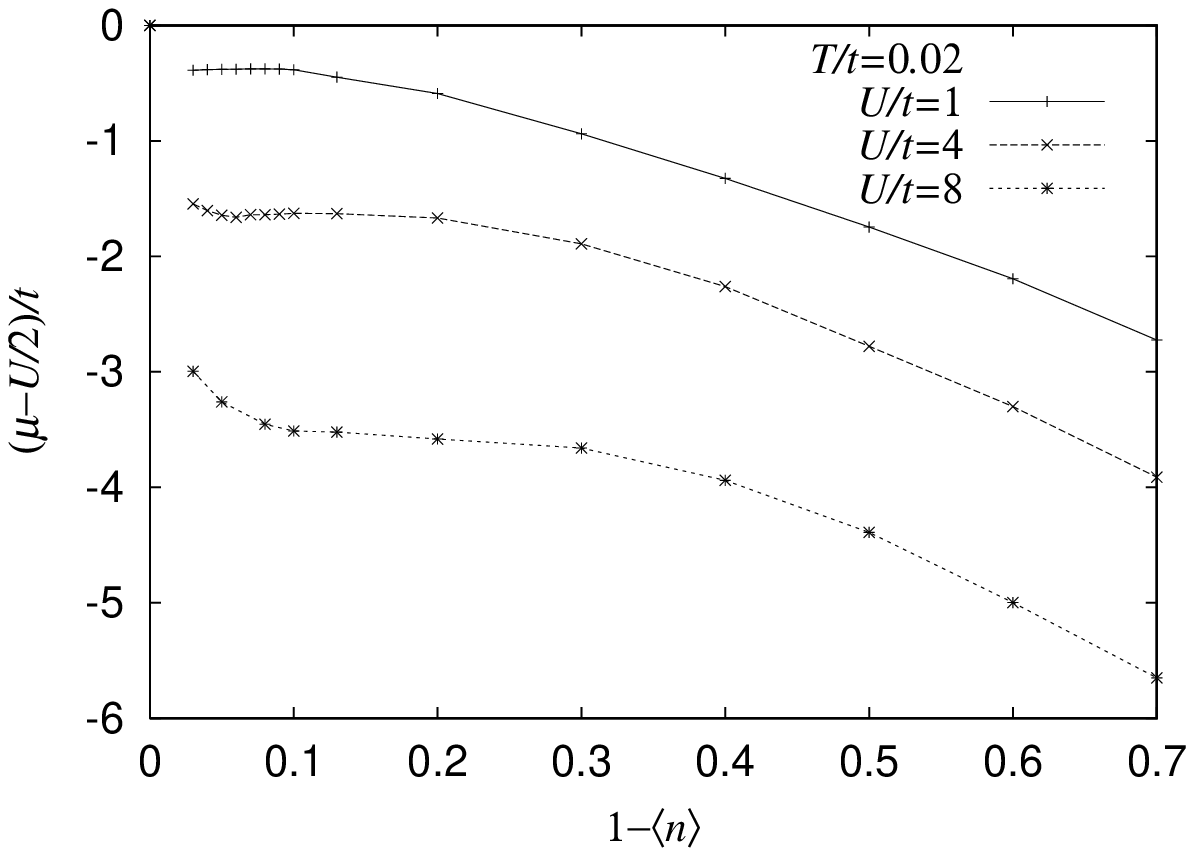}$$
\end{center}
\caption{$\mu-U/2$ versus the hole concentration $1-\ave{n}$. Near half filling, $\kappa=\del\ave{n}/\del\mu$ appears to diverge as $\ave{n}\to1$. For $U/t=4$ and $8$, the chemical potential shows a strong $\ave{n}$ dependence for $\ave{n}\ge 0.95$ and $0.9$, respectively. This occurs because slight but finite spectral weights still remain inside the $T=0$ Mott gap of order of $U$ due to thermal effects. Because of the smaller residual states for larger $U$, $\mu$ dependence of $\ave{n}$ increases with increasing $U$.}
\label{fig:ef-elt0U1,4,8}
\end{figure}

The main peak at every momentum on the magnetic Brillouin zone boundary is located below the Fermi level for $U/t=8$ and $\ave{n}=0.87$ even at $T/t=0.02$. This gives a larger Fermi volume than the Luttinger volume. Results on the momentum distributions also reveal these features~\cite{OPM}. Here, we note that the obtained Fermi volume decreases as $T\to0$. We also note that the present method gives a Fermi volume consistent with the Luttinger volume for $U/t=1$ and $\ave{n}=0.87$ at $T/t=0.02$. Suppressions of the Fermi degeneracy and thus $T_{\rm F}$ due to strong local repulsion in the proximity of the Mott insulator may produce the inconsistency with the Luttinger theorem for $U/t=8$ and $\ave{n}=0.87$. Actually, QMC calculations show that for $T/t\sim0.5$, the $(\pi,0)$ level is below the Fermi level~\cite{BulutScalapinoWhite94}. The strongly flat dispersion at low energies makes it difficult to speculate the $T=0$ properties from the present results.

\begin{figure}[htb]
\begin{center}\leavevmode
\epsfysize=5.5cm
$$\epsffile{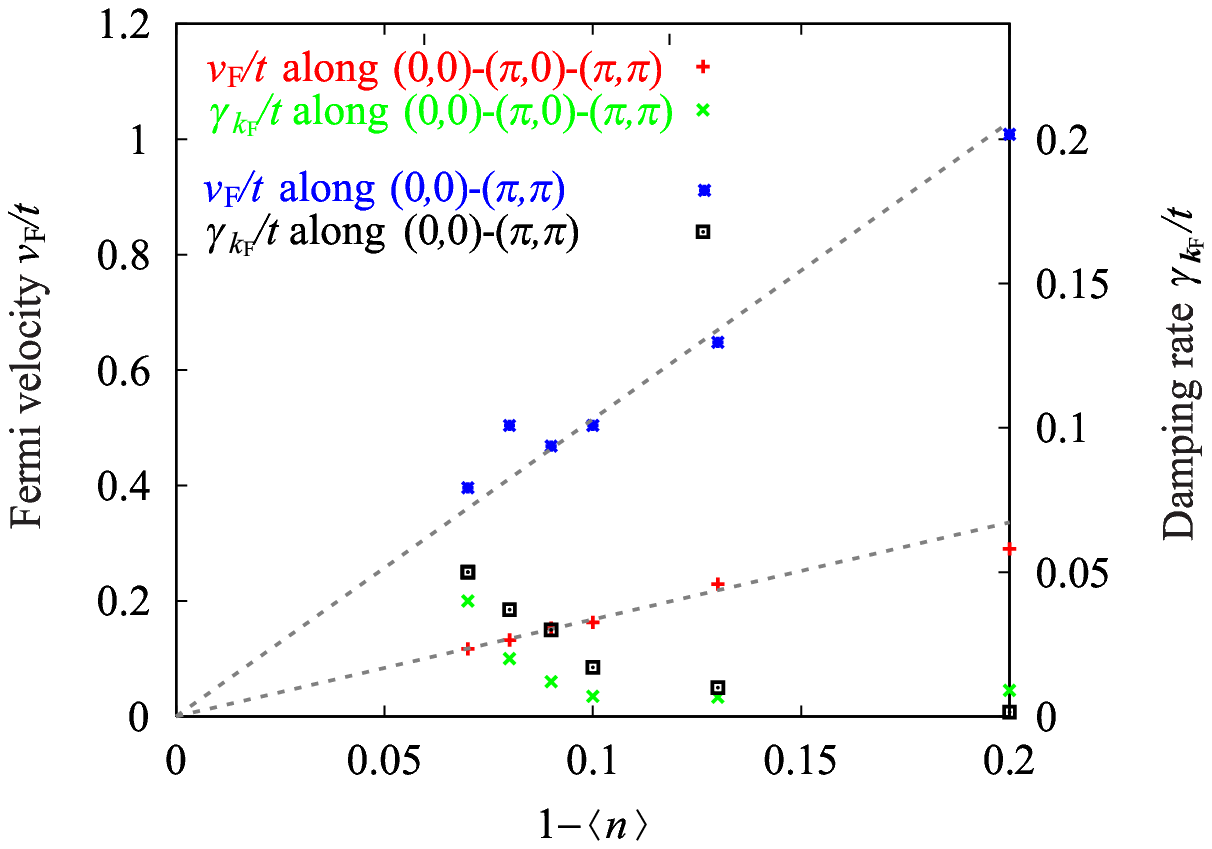}$$
\end{center}
\caption{Fermi velocities $v_{\rm F}=[|\nabla\veps^*(\mib{k})|]_{{\mibs k}={\mibs k}_{\rm F}}$ and damping rates $\gamma_{{\mibs k}_{\rm F}}$ as a function of $1-\ave{n}$ at Fermi momenta $\mib{k}_{\rm F}$. $\gamma_{{\mibs k}_{\rm F}}$ is determined as the width of the lorentzian fit to $A(\omega,\mib{k})$ at $T/t=0.02$. In both momentum directions, the quasiparticle mass diverges towards the MIT. However, $v_{{\mibs k}_{\rm F}}$ around $(\pi,0)$ and $(0,\pi)$ is less than that along $(0,0)$-$(\pi,\pi)$. The lines are guide to the eye. For $0.94\le\ave{n}\le1$, low-energy excitations in $A(\omega,{\mib k})$ exhibit a too broad structure to recognize quasiparticles. Therefore, the Fermi surface is also ill-defined there.}
\label{fig:quasit0U4}
\end{figure}

Figure~\ref{fig:quasit0U4} shows the mass divergence around both momenta $(\pi,0)$ and $(\pi/2,\pi/2)$, although the anisotropy is large. The diverging mass itself is consistent with the Brinkman-Rice picture~\cite{BrinkmanRice70}, reflecting the generic suppression of the Fermi degeneracy near the MIT. However, the enhanced density of states due to the low-energy band is pinned at the top of the lower Hubbard band. This is the origin of new physics not contained in the Brinkman-Rice picture. Furthermore, damping rates $\gamma_{\mibs{k}_{\rm F}}$ exhibit a prominent increase as $\ave{n}\to1$. Therefore, towards the MIT, quasiparticles become ill-defined at least at $T=0.02t$, since $\gamma_{\mibs{k}_{\rm F}}\sim v_{\rm F}$.

We summarize a unified view of the MIT in the 2D Hubbard model beyond the Mott-Hubbard, Brinkman-Rice and Slater pictures:

Towards the MIT, the local spin moments develops. It increases with increasing $U$ through the reduction of the double occupancy. This produces the splitting of the upper and lower Hubbard bands as in the Mott-Hubbard picture~\cite{Mott,Hubbard}. At $T<J$, the large local moments tend to align antiferromagnetically to gain a superexchange energy, as in the Slater picture. This contributes to formation of the SDW-like dispersions in the Hubbard bands through the momentum dependence of the self-energy part. At low temperatures, a narrow band appears at the top of the lower Hubbard band for hole-doped cases. However, the origin of this narrow band contrasts with the Brinkman-Rice theory~\cite{BrinkmanRice70} and the DMFT, because the enhanced density of states already exists at the gap edge in the Mott insulator.  Since the dispersions in the Hubbard bands take the form of SDW-like dispersions, the level of the narrow band is located near the magnetic Brillouin zone in contrast with the Brinkman-Rice picture. The velocity of these low-energy excitations increases with increasing holes. But the narrow band is pinned at the top of the lower Hubbard band, in contrast with the results of the DMFT. These lead to the diverging charge compressibility.

\section{Comparison with high-$\Tc$ cuprates}
\label{sec:ARPES}

In \BSCCO{2}{2}{}{2}{8+\delta}, a strongly weak dispersion both around the $(\pi,0)$ and $(0,\pi)$ and along $(\pi,0)$--$(\pi,\pi)$ has been observed with the angle-resolved photoemission spectroscopy (ARPES)~\cite{ARPES,Ding_bilayer}. The reduction from the band-calculation data is significant. This flatness depends only weakly on both $T$ and the hole concentration; it persists below and above $T_{\rm c}$. This suppression of the dispersion is attributed to strong correlation effects related to the MIT, as discussed previously~\cite{BulutScalapinoWhite94,Assaad}. The OPM gives an intuitive explanation of such enhancement of the electron mass: Local AF or singlet correlations increase $|\tilde{t}_{\mibs{x},\mibs{x}'}|$ between the nearest-neighbor sets of $\mibs{x}$ and $\mibs{x}'$. This affects the dispersion in the level of the two-pole approximation so that the $(\pi,\pi)$ level in the lower Hubbard band shifts downwards and the dispersion around the $(\pi,0)$ and $(0,\pi)$ momenta becomes flatter. Previous analytic theories do not reproduce such strong correlation effects. The similarity of the dispersions between our results and high-$\Tc$ cuprates indicates that to capture essential features of the superconductivity, it is crucial to describe the correlation effects with strongly momentum-dependent renormalization shared by the Hubbard model and the cuprates near the MIT. We speculate that the low-energy narrow band plays an important role in recovering the quantum-mechanical coherence.

\section{Summary}
\label{sec:Summary}

The 2D Hubbard model at and near half filling has been studied by the OPM. This is a non-perturbative analytic theoretical framework improving the Hartree-Fock theory, the perturbation and one-loop theories of the self-energy part and the Hubbard and two-pole approximations~\cite{OPM}. The OPM captures crucial ingredients of the MIT derived from the Mott-Hubbard, Brinkman-Rice and Slater pictures together with the strong momentum dependence of the single-particle spectra, which emerges as a crucial aspect of the strong correlation and a source of {\it electron differentiation in the momentum space} nemurically obtained and also observed in high-$\Tc$ cuprates.
Further studies on the low-energy hierarchy are necessary to discuss the Fermi-liquid properties and the mechanism of the high-$\Tc$ cuprate superconductivity.

We thank F. F. Assaad for valuable discussions and providing results of Monte-Carlo calculations. The work was supported by the Japan Society for the Promotion of Science under grant number JSPS-RFTF97P01103.

\end{document}